\documentclass[3p,times,procedia]{elsarticle}
\usepackage{nupha_ecrc}

\volume{00}

\firstpage{1}

\journalname{Nuclear Physics A}

\runauth{Koichi Hattori and Yi Yin}

\jid{nupha}

\jnltitlelogo{Nuclear Physics A}

\usepackage{graphicx}
\usepackage{latexsym}
\usepackage{amsfonts}
\usepackage{amssymb}
\usepackage{amsmath}
\usepackage{bm}
\usepackage{mathrsfs}
\usepackage{slashed}

\usepackage{dcolumn}

\def \be {\begin{equation} }
\def \ee {\end{equation}}
\def \bea {\begin{eqnarray}}
\def \eea {\end{eqnarray}}
\def \bem {\begin{multline}}
\def \eem {\end{multline}}

\def \bes {\begin{subequations} }
\def \ees {\end{subequations}}

\def \pd {\partial}

\def \e {\epsilon}

\def \x {\xi}

\def \<{\langle}
\def \>{\rangle}
\def \+{\dagger}
\def \({\left(}
\def \){\right)}

\def \[{\left[}
\def \]{\right]}

\def \vB {\bm{B}}

\def \he {\hat{e}}

\newcommand{\bo}{{\bm{\omega}}}
\newcommand{\bB}{{\bm{B}}}
\newcommand{\bS}{{\bm{S}}}

\newcommand{\Bhat}{\hat {\bm{B}}}

\newcommand{\sgn}{ {\rm sgn} }

\def \vB {\bm{B}}

\usepackage[normalem]{ulem}  

\renewcommand\sout{\bgroup \color{red} \ULdepth=-.5ex \ULset}

\usepackage[bookmarks=true,colorlinks,linkcolor=blue,urlcolor=green,citecolor=red]{hyperref}

\usepackage[figuresright]{rotating}

\begin{document}

\begin{frontmatter}



\dochead{XXVIth International Conference on Ultrarelativistic Nucleus-Nucleus Collisions\\ (Quark Matter 2017)}

\title{Charge redistribution from novel magneto-vorticity coupling in anomalous hydrodynamics}


\author[Fudan]{Koichi Hattori}
\ead{koichi.hattori@riken.jp}

\author[MIT]{Yi Yin}
\ead{yiyin3@mit.edu}

\address[Fudan]{Physics Department and Center for Particle Physics and Field Theory, Fudan University, Shanghai 200433, China.}
\address[MIT]{Center for Theoretical Physics, Massachusetts Institute of Technology, Cambridge, MA 02139, USA.}

\begin{abstract}
We discuss new transport phenomena in the presence of both a strong magnetic field and a vortex field. 
Their interplay induces a charge distribution and a current along the magnetic field. 
We show that the associated transport coefficients can be obtained from 
a simple analysis of the single-particle distribution functions and also from the Kubo formula calculation. 
The consistent results from these analyses suggest that the transport coefficients 
are tied to the chiral anomaly in the (1+1) dimension 
because of the dimensional reduction in the lowest Landau levels. 
\end{abstract}


\end{frontmatter}

\section{Introduction}

It has been known that spectra of charged fermions are subject to the Landau level discretization in magnetic fields, 
and that the (1+1)-dimensional chiral fermions emerges in the lowest Landau level (LLL). 
This fact inspired a lot of new ideas in the last decade: 
To name a few, the interplay between the chiral anomaly and the magnetic field 
induces currents along the magnetic field which are now known as chiral magnetic effect and chiral separation effect 
(see, e.g., Refs.~\cite{Kharzeev:2013ffa, Kharzeev:2015znc, Huang:2015oca, Hattori:2016emy} for reviews). 
Moreover, the vortex in the chiral fluid, composed of massless fermions, 
also induces the currents along the vortex line, i.e., chiral vortical effects. 
They are intimately connected to the topological aspect of the chiral anomaly. 
The ultrarelativistic heavy-ion collisions contain all these ingredients, 
strong magnetic fields, vortices of the quark-gluon plasma (QGP), 
and topological structures of QCD vacuum. 
Recent progresses in the experimental search of these phenomena by Relativistic Heavy Ion Collider 
and Large Hadron Collider induced a lot of discussions in Quark Matter 2017 \cite{QM2017}. 
 
\begin{figure}[t]
     \begin{center}
              \includegraphics[width=0.5\hsize]{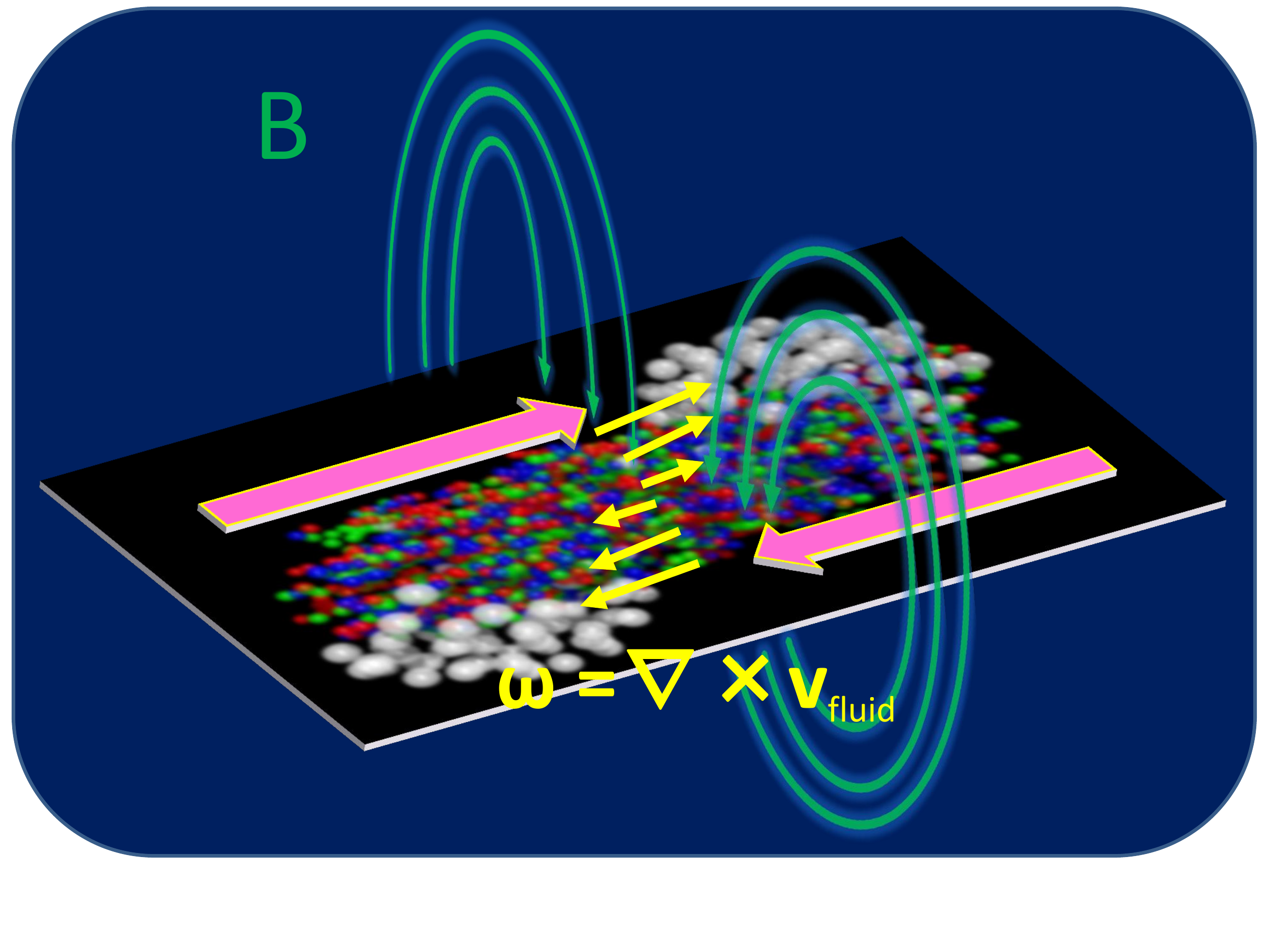}
     \end{center}
\vspace{-.8cm}
\caption{Strong magnetic field and vortical field induced by the ultrarelativistic heavy-ion collisions.}
\label{fig:HIC}
\end{figure}

In our contribution, we have proposed that an interplay between 
effects of the strong magnetic field and the vortices in QGP (cf., Fig.~\ref{fig:HIC}) gives rise to 
novel transport phenomena, based on our recent publication~\cite{Hattori:2016njk}. 
While there have been numerous studies on the separate effects 
of the magnetic field and the vortex field, their interplay has not been explored. 
For example, one finds that, in the pioneering work on the anomalous hydrodynamics~\cite{Son:2009tf}, 
the vorticity and magnetic field are treated as the first order in the gradient expansion, 
and their coupling is dropped as a higher-order effect. 
However, since the medium does not give rise to a screening effect on the magnetic field, 
one needs to go beyond the conventional gradient expansion in the regime of magnetohydrodynamics. 
In such a regime, the interplay between the magnetic and vortical fields becomes important. 
We showed that the interplay induces the redistribution of vector charges and an axial current. 
Especially, based on field-theoretical analysis by the Kubo formula, 
we explicitly showed that the associated transport coefficients 
are tied to the chiral anomaly in the (1+1) dimensions. 

\section{Consequences of the magneto-vorticity coupling}

When the fluid  has a nonvanishing vorticity $  \bo$, 
it is coupled to the intrinsic angular momentum of a fermion, i.e., the spin $ {\bm S} $. 
Therefore, the distribution function in a local equilibrium is shifted as $f_{0}(\e+\Delta \e)$ 
with $f_{0}$ being the global equilibrium distribution function. 
The energy shift is given by the coupling $\Delta \e = - {\bm S} \cdot \bo$. 
This effect has been considered for the coupling between the quarks and the vortex in QGP~\cite{Becattini:2013fla}, 
which led to the idea of the $ \Lambda $-polarization measurement. 

We consider what would happen in the presence of the strong magnetic field induced in the heavy-ion collisions. 
First of all, the strong magnetic field confine the fermions into the lowest Landau levels. 
Also, the large magnitude of the Zeeman splitting selects out the unique ground state 
with the energetically favored spin configuration. 
The higher states are all gapped with the large interval $\sim \sqrt{2 q_f B}  $. 
Therefore, considering the ground state, the spin direction is 
determined by the direction of the magnetic field as $\bS_{R/L} = \frac{1}{2}  \sgn(q_f) \Bhat$, 
which is independent of the chirality. 
Consequently, the sign of the energy shift is also determined by the direction of the magnetic field as 
\begin{equation}
\label{e-omega}
\Delta \epsilon_{LLL}^{\pm} =\mp \frac{1}{2} \sgn(q_{f})\, \Bhat \cdot \bo
\, , 
\end{equation}
where the upper and lower signs are for a particle and an antiparticle, respectively. 
Below, we take $\vB=B\, \hat{e}_{3}$ without loss of generality. 

We now compute the density and current of the chiral fermions in the LLL. 
We shall first look at the density. 
Expanding $f_{0}(\e')$ with respect to $\Delta \e$ for the perturbative vorticity, 
and using the linear dispersion relation of the right-handed LLL fermion, $\e_{\rm LLL}= + p^{3}$, 
we find the density of the right-handed fermion as 
\begin{eqnarray}
\label{Delta-nR}
\Delta n_{R}&=&
\frac{|q_{f}\, B|}{2\pi}
\left[\, \Delta \e_{LLL}^{+} \int^{\infty}_{0}\, \frac{dp^{3}}{2\pi}\, 
\frac{\pd\, f_{0}(p^{3})}{\pd p^{3}}
 +
 \Delta \e_{LLL}^{-}
\int^{0}_{-\infty}\, \frac{dp^{3}}{2\pi}\, 
\frac{\pd\, \bar{f}_{0}(p^{3})}{\pd p^{3}}
\, \right ]\, 
\nonumber \\
&=& q_f \frac{C_{A} }{4} ( \bB \cdot \bo ) \, \[\,  f_{0}(0)+\bar{f}_{0}(0)\, \]
\nonumber\\
&=& q_f \frac{C_{A} }{4} (\bB \cdot \bo )
\, .
\end{eqnarray}
Here, the factor of $|q_{f}B|/(2\pi)$ is the density of states in the LLL per unit transverse area, 
and the Fermi-Dirac distribution functions of particles and antiparticles are given by 
$f_{0}(\e)= 1/[e^{(\e-\mu)/T}+1] $ and $ \bar{f}_{0}(\e)= 1/[e^{-(\e-\mu)/T}+1]$, respectively. 
We used a remarkable fact that the sum of the particle and antiparticle distributions at the origin 
is identically equal to unity, $f_{0}(0)+\bar{f}_{0}(0)=1$, 
meaning that the density shift is independence of temperature $T$ and chemical potential $\mu$.
Since the spin direction, and thus the energy shift, is independent of the chirality, 
we find the same result for the left-handed fermions as $\Delta  n_{L}=\Delta n_{R}$. 
In the basis of the vector and axial-vector channels, 
we find 
\begin{eqnarray}
\Delta n_{V}
= q_f  \frac{ C_{A} }{2}  (\bB \cdot \bo ) \, , \ \ \ \ 
\Delta n_{A}
= 0
\, ,
\label{jV}
\end{eqnarray}
with $  C_A = 1/(2\pi^2)$. An additional factor of $  1/2$ is originated from the spin size. 

\begin{figure}[t]
     \begin{center}
              \includegraphics[width=0.55\hsize]{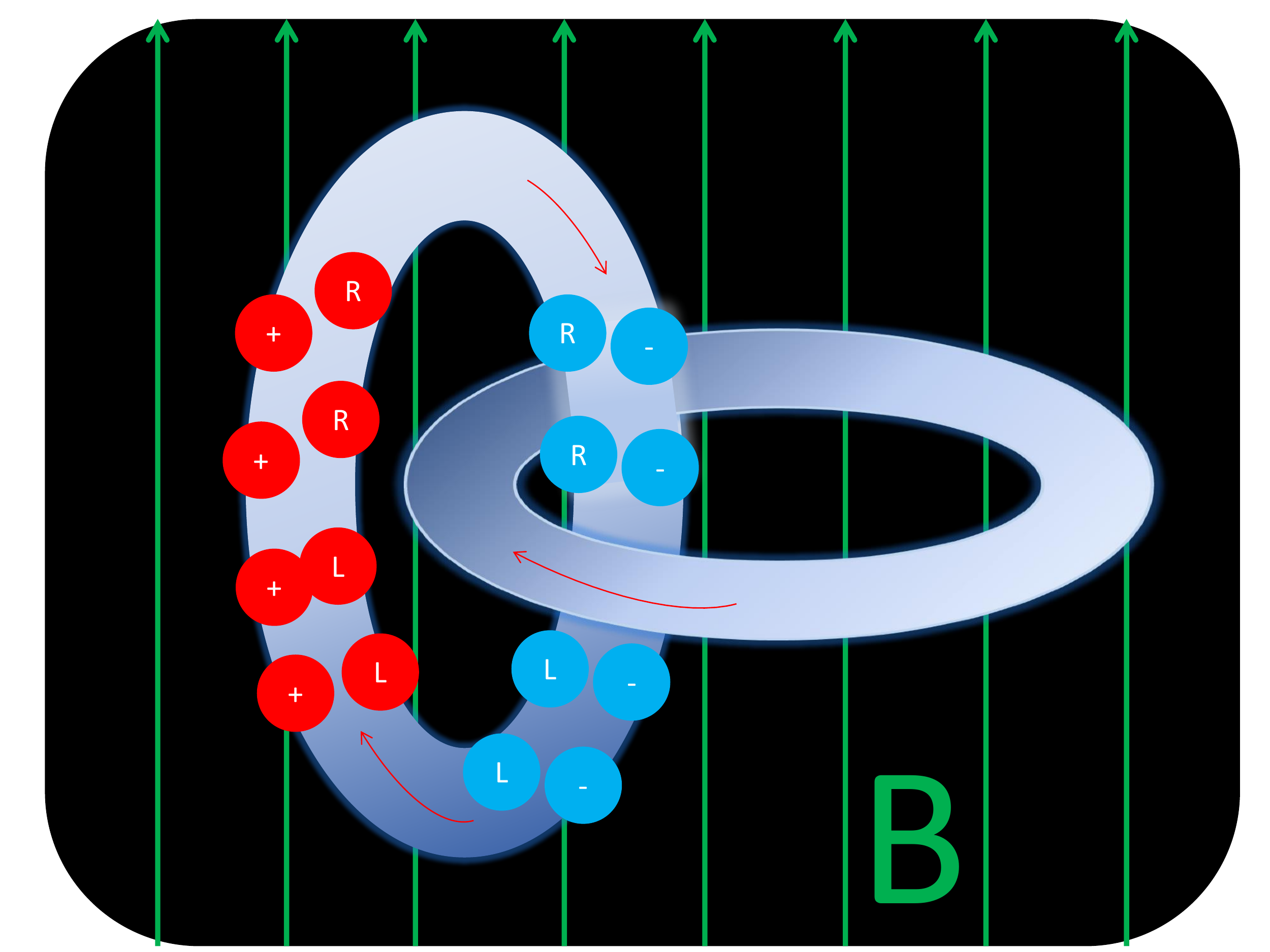}
     \end{center}
\vspace{-.5cm}
\caption{Vector-charge distribution and induced axial current 
along the parallel configuration of the magnetic and vortex lines.}
\label{fig:magneto-vortex}
\end{figure}

Next, one can perform a similar computation for the current 
composed of the LLL chiral fermions moving along $\he_{3}$ with the speed of light. 
The direction of the velocity, and thus of the current, is uniquely determined 
by specifying the spin direction and the chirality. Therefore, we have 
$\Delta j^{3}_{R}= \sgn(S_R) \Delta n_{R} $ and $ \Delta j^{3}_{L}= -  \sgn(S_L) \Delta n_{L}$, 
where the minus sign in the $ \Delta j^3_L $ is originated from the velocity in the negative third direction 
and the spin directions of the LLL fermions are aligned along the magnetic field. 
We then find the shifts of the vector and axial-vector currents as  
\begin{eqnarray}
\Delta j^{3}_{V} 
= 0 \, , \ \ \ \ 
\Delta j^{3}_{A} 
= \vert q_f \vert \frac{  C_{A} }{2}  (\bB \cdot \bo ) \Bhat
\label{jA}
\, .
\end{eqnarray}
The above heuristic derivation tells us the induced vector-charge density 
and the axial current along the magnetic field. 

The above findings are schematically shown in Fig.~\ref{fig:magneto-vortex}. 
When the vortex field has an parallel component to the background magnetic field, 
the vector charge is accumulated along the vortex line. There, the axial current is also induced. 
The sign of the charge and the direction of the current are determined by 
the relative direction of the vortex and magnetic fields.

\begin{figure}[t]
     \begin{center}
              \includegraphics[width=\hsize]{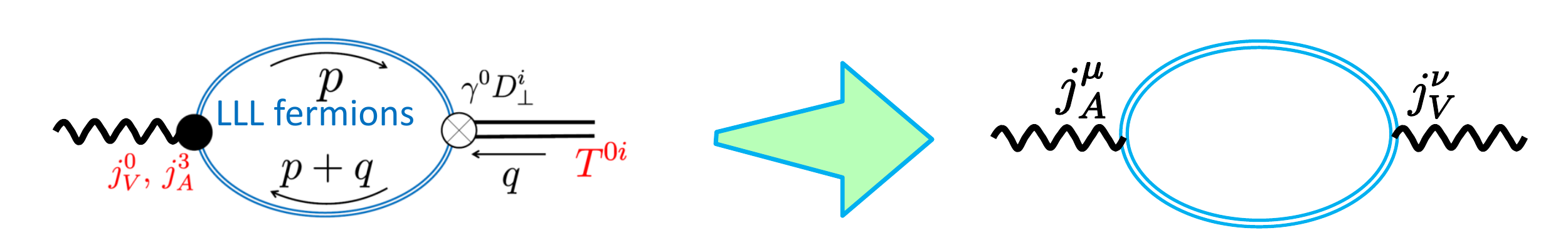}
     \end{center}
\vspace{-.5cm}
\caption{Linear-response diagram relating the purturbation by the vortex to the induced current, 
which is tied to the anomaly diagram in the (1+1) dimensions.}
\label{fig:Kubo}
\end{figure}

Looking again at the above results in Eqs.~(\ref{jV}) and (\ref{jA}), 
we notice that the induced vector-charge density and the axial current are 
proportional to the anomaly coefficient $ C_A $. 
This is not an accidental coincidence. 
In Ref.~\cite{Hattori:2016njk}, we have shown that those new transport phenomena 
are indeed tied to the chiral anomaly in the (1+1) dimension 
on the basis of the Kubo formula which is diagrammatically shown in Fig.~\ref{fig:Kubo}. 
The linear response is captured by a retarded correlator between 
the energy-momentum tensor perturbed by the vortex field and the induced current. 
The LLL fermion propagators have the (1+1)-dimensional form (see Ref.~\cite{Hattori:2016njk} for more details). 
The key observation is that one can establish an explicit relation between 
the retarded correlator and the anomaly diagram in the (1+1) dimension, that is, 
the two-point function between the vector and axial-vector currents 
(which is also connected to the vector-current correlator via a simple relation). 
From the well-known results of the Schwinger model, 
we confirm all of the results in Eqs.~(\ref{jV}) and (\ref{jA}) on the basis of the linear response theory. 
Especially, we naturally find that the transport coefficients are proportional 
to the anomaly coefficient $ C_A $, and are independent of temperature and density. 
These interesting properties are attributed to the properties of the anomaly diagram for the massless fermions 
(see, e.g., Ref.~\cite{FHYY} for the explicit computation of the current correlators). 

\section{Summary and prospects} 
We discussed new transport phenomena induced by the magneto-vorticity coupling. 
Our analyses by the Kubo formula suggest that they are tied to the chiral anomaly in (1+1) dimensions. 
We also briefly discussed that the coupling between the magnetic and vortex fields 
induces a new hydrodynamic instability in the presence of a finite axial chemical potential. 
This instability amplifies the fluid velocity exponentially in time. 
More detaled account will appear elsewhere \cite{instability}. 

\vspace{0.3cm}

\section*{Acknowledgement}
The research of KH is supported by China Postdoctoral Science Foundation under Grant No. 2016M590312. 
This material is based upon work supported in part by the U.S. Department of Energy, Office of Science, Office of Nuclear Physics, under Contract Number de-sc0011090 (YY) and within the framework of the Beam Energy Scan Theory (BEST) Topical Collaboration.


\bibliographystyle{elsarticle-num}
\bibliography{bib}


\end{document}